\documentclass[12pt]{article}
\RequirePackage{ifpdf}
\ifpdf\RequirePackage[raiselinks=false,colorlinks=true,citecolor=blue,urlcolor=blue,linkcolor=blue,bookmarksopen=true,pdftex]{hyperref}\else
\RequirePackage[raiselinks=false,colorlinks=true,citecolor=blue,urlcolor=blue,linkcolor=blue,bookmarksopen=true,dvips]{hyperref}\fi
\usepackage[latin1]{inputenc} 
\usepackage[T1]{fontenc} 
\usepackage{amsfonts,amsmath}
\usepackage{latexsym}
\usepackage{color}  %
\usepackage{pst-node} 
\usepackage{ae,aecompl,amsbsy,amssymb}
\newcommand{\huffUrl}{ftp://ftp-sop.inria.fr/lemme/Laurent.Thery/TwoSquares}
\newtheorem{theo}{Theorem}[section]
\newtheorem{Def}{Definition}[section]

\newcommand{\imod}[2]{#1\, \hbox{ \tt mod}\, #2}
\renewcommand{\mod}[3]{#1\, \equiv\, #2\, [#3]}
\renewcommand{\div}[2]{#1\, |\, #2}
\newcommand{\incl}[2]{#1 \subset #2}
\newcommand{\size}[1]{|#1|}
\newcommand{\order}[1]{o(#1)}
\newcommand{\idiv}[2]{#1/#2}
\newcommand{\zn}[1]{{(\mathbb{Z}/#1\mathbb{Z})}^{*}}
\renewcommand{\prime}[1]{{\it prime}(#1)}
\newcommand{\coprime}[2]{{\it coprime}(#1,#2)}
\renewcommand{\gcd}[2]{\hbox{$#1$}\,\hbox{$\hat{\,\,}$}\,\hbox{$#2$}}

\title{Primality Tests and Prime Certificate}
\author{Laurent Th{\'e}ry \\
Marelle Project - INRIA Sophia Antipolis
}
\date{}
\begin{document}
\maketitle
\begin{abstract}
This note presents a formalisation done in {\sc Coq} of  Lucas-Lehmer test and
Pocklington certificate for prime numbers. They both are direct
consequences of Fermat little theorem. Fermat little theorem is proved
using elementary group theory and in particular Lagrange theorem.

\end{abstract}
\section{Definitions and Notations}

In order to present our formalsation, we first need to introduce
some functions and predicates over natural numbers, lists and sets.

\subsection{Natural numbers}

The predicates over the natural numbers are the following:
\begin{itemize}
\item[-] {\bf Divisibility}: $n$ divides $m$, written $\div{n}{m}$, if there exists 
a number $q$ such that $m=nq$.
\item[-] {\bf Primality}: $p$ is prime, written $\prime{p}$, if $p$ has
{\it exactly} two positive divisors $1$ and $p$.
\item[-] {\bf CoPrimality}: $p$ and $q$ are co-prime, written $\coprime{p}{q}$, if $1$ is their unique positive common divisor.
\item[-] {\bf Modulo}: $p$ is equal to $q$ modulo $n$, written $\mod{p}{q}{n}$, if $n$ divides $p-q$.
\end{itemize}
and the functions are:
\begin{itemize}
\item[-] {\bf Gcd}: the greatest common divisor of two numbers $p$ and $q$ is
written $\gcd{p}{q}$.
\item[-] {\bf Quotient}: the integer quotient of the division of $p$ by $q$ is written $\idiv{p}{q}$. 
\item[-] {\bf Remainder}: the remainder of the division of $p$ by $q$ is 
written $\imod{p}{q}$.
\item[-] {\bf Euler function}: $\Phi(n) = \sum_{i=1}^{n-1}{({\tt if}\,\, \coprime{i}{n}\,\, {\tt then}\,\, 1\,\, {\tt else}\,\, 0)}$.
\end{itemize}
 \subsection{Lists}
Lists are denoted as $[ a_1, a_2, \dots, a_n]$. 
We write the size of a list $L$ as $\size{L}$,
the concatenation of two lists $L_1$, $L_2 $ as $L_1 + L_2$ and
the fact that an element $a$ belongs to a list $L$ as $a \in L$.
 \subsection{Sets}
Sets are denoted as $\{ a_1, a_2, \dots, a_n\}$. We write
the size of a set $G$ as $\size{G}$ , the fact that an element 
$a$ belongs to a set $G$ as $a \in G$, the fact that a set $G_1$
is included in a set $G_2$ as $G_1 \subset G_2$.
Over sets, we define the notions of finite monoid and finite group:
\begin{itemize}
\item[-] {\bf Finite Monoid}: $(G, *)$ is a finite monoid, iff
\begin{itemize}
\item[] $G$ is finite: $G = \{e, a_1, a_2, ..., a_n\}$,
\item[]the operation $*$ is internal: 
$\hbox{if} \, a \in G\, \hbox{and} \, b \in G\, \hbox{then} \, ab\in G$,
\item[]the operation $*$ is associative: $a(bc) = (ab)c$,
\item[]the element $e$ is neutral: $ea = a = ae$.
\end{itemize}
\item[-] {\bf Finite Group}: $(G, *)$ is a finite group, iff
\begin{itemize}
\item[] $G$ is finite: $G = \{e, a_1, a_2, ..., a_n\}$,
\item[]the operation $*$ is internal: 
$\hbox{if} \, a \in G\, \hbox{and} \, b \in G\, \hbox{then} \, ab\in G$,
\item[]the operation $*$ is associative: $a(bc) = (ab)c$,
\item[]the element $e$ is neutral: $ea = a = ae$,
\item[]every element has an inverse: $aa^{-1} = e = a^{-1}a$.
\end{itemize}
\end{itemize}
A group $(H, *)$ is a subgroup of a group $(G, *)$ if $G\subset H$.
\section{Basic Theorems}

Four basic theorems are mainly needed for our development: Gauss theorem for division, Bezout theorem for gcd,
the fact that$\Phi(p)= p-1$ for $p$ prime and Lagrange for the cardinality of subgroup.

\begin{theo}[Gauss]
  \label{theo/gauss}
If $\div{m}{np}$ and $\coprime{m}{n}$ then $\div{m}{p}$.
\end{theo}

This theorem does not belong to our development, 
nevertheless we outline
its proof. The key point of the proof is
that divisibility is compatible with the substraction:
if $\div{m}{n}$ and $\div{m}{p}$ then $\div{m}{n-p}$.
Now, we have the hypothesis $\div{m}{np}$ and  we also have
that $\div{m}{mp}$.
Remembering Euclid algorithm and using the compatibility of
the substraction we can derive that $\div{m}{(\gcd{m}{n})p}$.
 As we have $\coprime{m}{n}$, we get the expected result $\div{m}{p}$.

\begin{theo}[Bezout]
  \label{theo/bezout}
Let $m$ and $n$ be two integers, then there exist $u$ and $v$ such
that $mu+nv=\gcd{m}{n}$.
\end{theo}
Once again the proof of this theorem follows Euclid algorithm
to compute the gcd of $m$ and $n$.

\begin{theo}
  \label{theo/phip}
Let $p$  be a prime number, $\Phi(p) = p -1$
\end{theo}
Since $p$ is prime, if $1 \le i < p$ then $\coprime{i}{p}$, so
$\Phi(p) = p - 1$.

\begin{theo}[Lagrange]
  \label{theo/lagrange}
  If $(H, *)$ is a subground of $(G, *)$ then $\div{\size{H}}{\size{G}}$.
\end{theo}
Let $H=\{e, a_1, a_2, \dots, a_n\}$ and $G=\{e, b_1, b_2, \dots, b_m\}$,
we have $\incl{H}{G}$. We build the increasing sequence $(L_i)_{i \le m}$
of lists as follows:
\begin{itemize}
\item[] $L_0 = [e, a_1, a_2 \dots a_n]$;
\item[] if ${b_{i+1}} \in {L_i}$, ${L_{i+1}=L_{i}}$;
\item[] if ${b_{i+1}} \not\in {L_i}$,
 ${L_{i+1}=[b_{i+1}e, b_{i+1}a_1, b_{i+1}a_2, \dots, b_{i+1}a_n] + L_{i}}$.
\end{itemize}
We use the convention that $a_0 = e$ and $b_0 = e$.
It is easy to show that for all $i \le m$ we have $\div{\size{H}}{\size{L_i}}$ and 
$b_i \in L_m$. We are left with proving that $\size{L_m} = \size{H_m}$.
To do so, we just need to show that the elements of $H$ occurs only
once in $L_m$. By contradiction, suppose $b_k$ occurs more than
once in $L_m$. There are two possibilities: either there exists $i$ such
that $b_k$ occurs more than once in $L_{i}$ but not in $L_{i-1}$\footnote{
This includes also the degenerated case where $b_k$ occurs twice
in $L_0$},
or $b_k$ occurs in $L_i$ and $L_{i} - L_{i-1}$. In the first case,
there exist $u$ and $v$, $b_{i}a_u = b_k = b_{i}a_v$. Simplifying
by $b_{i}^{-1}$, we get $a_u = a_v$. So as $G$ is a set,
we have $u = v$. This contradicts the fact that $b_k$ occurs more
than once in $L_i$. In the second
case, there exist $u$, $v$ and $j$ with $j < i$ such
that  $b_{i}a_u = b_k = b_{j}a_v$. Simplifying by $a_u^{-1}$,
we get $b_i = b_{j}(a_va_u^{-1})$. As $a_v \in G$ and  $a_u \in G$,
there exists a $l$ such that $b_i = b_{j}a_l$, so $b_i \in L_j$.
This contradicts the fact that $b_i \not\in L_{i-1}$ that is true
by construction since $L_i \neq L_{i -1}$. 

\section{Group of invertible elements}

From a monoid we can extract a group by taking its invertible elements.
This section explicits how this group is constructed and states some basic properties.

\begin{Def}
  \label{def/inv}
  Let $(G, *)$ be a finite monoid, we define $I(G)$ as $\{a \in G\, |\, \exists c\in G, ca = e = ac\}$.
\end{Def}

\begin{theo}
  \label{theo/inv}
Let $(G, *)$ be a finite monoid, $(I(G), *)$ is a finite subgroup.
\end{theo}
$I(G)$ is finite since $I(G) \subset G$. The operation is internal
since if $a$ and $b$ are in $I(G)$, then there exist $c$ and $d$
such that $ac=e=ca$ and $bd = e = db$. It follows that
$(ab)(dc) = e = (dc)(ab)$ so $ab \in I(G)$. The operative is 
associative since $(G, *)$ is a monoid. We have $ee = e= ee$,
so $e \in I(G)$ and as it is a neutral element in $G$, it is
also a neutral element in $I(G)$. Every element has an inverse
by construction.

\begin{Def}
  \label{def/znzm}
  Given $n$, we define $\mathbb{Z}/n\mathbb{Z}$ as $\{ i\, |\, 0 \le i < n\}$ .
\end{Def}

\begin{Def}
  \label{def/mmod}
  Given  $n$, we define the operation $\otimes$ as
$a \otimes b = \imod{(ab)}{n}$.
\end{Def}

\begin{theo}
  \label{theo/znzmonoid}
Given $n$, $(\mathbb{Z}/n\mathbb{Z}, \otimes)$ is a finite monoid.
\end{theo}
$\mathbb{Z}/n\mathbb{Z}$ is finite. The operation $\otimes$
is internal since $0 \le \imod{a}{n} < n$.
The operation is associative since $a \otimes (b \otimes c) = \imod{(abc)}{n}
= (a \otimes b) \otimes c$. 1 is a neutral element. Note that operation
$\otimes$ is also commutative.

\begin{Def}
  \label{def/znz}
  Given $n$, we define $\zn{n}$ as $I(\mathbb{Z}/n\mathbb{Z})$.
\end{Def}

\begin{theo}
  \label{theo/znzgroup}
Given $n$, $(\zn{n}, \otimes)$ is a finite group.
\end{theo}
This is a direct consequence of Theorems~\ref{theo/znzmonoid}~and~\ref{theo/inv}.

\begin{theo}
  \label{theo/znzsize}
Given a number $n$,  $ \size{\zn{n}} = \Phi(n)$.
\end{theo}
Let $a \in \zn{n}$, so there exists $c$ such that $a \otimes c = 1$.
So $\imod{(ac)}{n}=1$, $\div{n}{ac - 1}$ and there exists a $d$ such that $ac - dn = 1$
so by Theorem~\ref{theo/bezout}, we have $\coprime{a}{n}$. Reciprocally
if $\coprime{a}{n}$ by Theorem~\ref{theo/bezout} there exist $u$ and $v$
such that $ua + vn = 1$, so $u\otimes a = 1$.

\begin{theo}
  \label{theo/zpzsize}
Given a prime number $p$,  $ \size{\zn{p}} = p - 1$.
\end{theo}
This is a direct consequence of Theorems~\ref{theo/phip}~and~\ref{theo/znzsize}.

\section{Order of an element}

Given an element $a$ of a group, we can construct a subgroup by repetitively
multiplying $a$ by itself. The cardinality of this subgroup is called the
{\it order} of the element. This section explicits this constructed and
state some basic properties. The last one is the famous
Fermat Little Theorem which is at the base of Pocklington certificate.

 \begin{Def}
  \label{def/ha}
  Let $(G, *)$ be a finite group and $a$ an element of $G$,
  we define $H_a = \{a^i | i \in \mathbb{N}\}$,
\end{Def}
Note that in the definition, we take as convention that $a^{0} = e$.
\begin{Def}
  \label{def/order}
  Let $(G, *)$ be a finite group and $a$ be an element of $G$,
  we define  $o(a)$, the order of the element $a$, as the smallest number such that
there exists $k < o(a)$ such that $a^k = a^{o(a)}$.
\end{Def}
First of all, $H_a$ is finite since $H_a \subset G$.
It follows there is a least one repetition in $[1, a, a^2, \dots a^{|G|}]$.
So the definition of $o(a)$ makes sense. 

\begin{theo}
  \label{theo/pow}
  Let $(G, *)$ be a finite group and $a$ be an element of $G$, we have
$a^{o(a)} = e$.
\end{theo}
There exists $k < o(a)$ such that $a^k = a^{o(a)}$. 
Multiplying  on both side by $a^{-k}$ we get $a^0 = a^{o(a)-k}$. Since
$o(a)$ was the smallest number for which there is a repetition, 
it implies that $k = 0$ and $a^{o(a)} = e$.

\begin{theo}
  \label{theo/powdiv}
  Let $(G, *)$ be a finite group, $a$ be an element of $G$ and $n$ be a number,
$a^{n} = e$ if and only if $\div{o(a)}{n}$.
\end{theo}
Suppose $a^{n} = e$,
by Definition~\ref{def/order}, we have $o(a) \le n$. Iteratively multiplying
by $a^{-o(a)}$ on both side of the equation $a^{n} = e$
we get $a^{\imod{n}{o(a)}} = e$.
Since ${\imod{n}{o(a)}} < o(a)$, it implies that $\imod{n}{o(a)} = 0$
so $\div{o(a)}{n}$. Conversely, suppose $\div{o(a)}{n}$, so there exists $k$
such that $n = ko(a)$. We have $a^{n} =a^{ko(a)} ={(a^{o(a)})}^k = e^k = e$. 

\begin{theo}
  \label{theo/dec}
  Let $(G, *)$ be a finite group and $a$ be an element of $G$,
$H_a = \{1, a, a^2, \dots, a^{o(a)-1}\}$  and $\size{H_a}=o(a)$.
\end{theo}
This is a direct consequence of Definition~\ref{def/ha} and
Theorem~\ref{theo/pow}.

\begin{theo}
  \label{theo/mult}
  Let $(G, *)$ be a finite group and $a$ an element of $G$,
$(H_a, *)$ is a finite subgroup  of $(G,*)$.
\end{theo}
$H_a$ is finite.
The operation $*$ is internal since $a^ia^j = a^{i+j}$.
The operation $*$ is associative since $a^i(a^ja^k) = a^{i+j+k}= (a^i a^j)a^k$.
Every element has an inverse  $a^ia^{o(a)-i} = e = a^{o(a)-i}a^i$.
Note that this group is also commutative.

\begin{theo}
  \label{theo/order}
  Let $(G, *)$ be a finite group and $a$ an element of $G$,
we have $\div{o(a)}{\size{G}}$.
\end{theo}
This is a direct consequence of Theorem~\ref{theo/lagrange}
and $\size{H_a}=o(a)$.

\begin{theo}
  \label{theo/orderzn}
  Let $n$ be a number and $a \in \zn{n}$, we have $\div{o(a)}{\Phi(n)}$.
\end{theo}
This is a direct consequence of Theorems   \ref{theo/order}~and~\ref{theo/znzsize}.

\begin{theo}
  \label{theo/orderzp}
  Let $p$ be a prime number and $a \in \zn{p}$, we have $\div{o(a)}{p-1}$.
\end{theo}
This is a direct consequence of Theorems \ref{theo/order}~and~\ref{theo/zpzsize}

\begin{theo}
  \label{theo/fermatg}
Let $n$ be a number and $\coprime{a}{n}$ then $\mod{a^{\Phi(n)}}{1}{p}$.
\end{theo}
As $\gcd{a}{b}=\gcd{(\imod{a}{b})}{b}$ and $\mod{a^i}{({\imod{a}{n}})^i}{n}$,
we can restrict ourselves to the case in which $a \in (\mathbb{Z}/n\mathbb{Z})^*$.
Using Theorem \ref{theo/orderzn}, we have $\div{\order{a}}{\Phi(n)}$.
By definition of the order,
it follows that $a^{\Phi(n)} \equiv a ^ {k\order{a}} \equiv (a ^{\order{a}})^k
\equiv 1 ^k \equiv 1\, [p]$ for some $k$.

\begin{theo}[Fermat Little Theorem]
  \label{theo/fermat}
  If $\prime{p}$ and $\coprime{a}{p}$ then $\mod{a^{p-1}}{1}{p}$.
\end{theo}
This is a direct consequence of Theorems~\ref{theo/phip}~and~\ref{theo/fermatg}.

\section{Lucas-Lehmer test\label{lucas}}

The previous sections have introduced all the material needed to present
Lucas-Lehmer test. This test gives a direct way of checking primality for Mersenne numbers.

 \begin{Def}
  \label{def/kn}
  Let $n$  be a number,
  we define $K_n$ as ${(\mathbb{Z}/n\mathbb{Z})}^{2}$, i.e. $K_n = \{(a, b) \,|\, 0 \le a \le n \,\hbox{and\,}\, 0 \le b \le n\}$.
\end{Def}
\begin{Def}
  \label{def/kplus}
  we define  the operation $\oplus$ as
 $(a_1,b_1) \oplus (a_2, b_2) = (\imod{(a_1 + b_1)}{n}, 
\imod{(b_1 + b_n)}{n})$.
\end{Def}
\begin{Def}
  \label{def/kmult}
  we define  the operation $\odot$ as
 $(a_1,b_1) \,\odot\, (a_2, b_2) = (\imod{(a_1a_2 + 3 b_1b_2)}{n}, 
\imod{(a_1b_2 + a_2b_1)}{n})$.
\end{Def}

\begin{Def}
  \label{def/kexp}
  we define the power as 
 $(a,b)^n = \underbrace{(a,b) \,\odot\, (a, b) \dots (a,b)}_{n}$.
\end{Def}

\begin{Def}
  \label{def/sn}
For $n> 1$,  we define two elements of $K_n$ $w$ as $(2,1)$, 
$v$ as $(2,n-1)$ and we define the sequence $(S_m)_{m \in \mathbb{N}}$ 
over the natural numbers such that  $S_0 = 4$ and $S_{m+1} = S_m^2 -2$.

\end{Def}

\begin{theo}
  \label{theo/wv}
For $n > 1$, we have $w \,\odot\, v = (1, 0)$,
\end{theo}
We have \begin{align*}
w \,\odot\, v 
&= (2, 1)\,\odot\, (2, n -1) \\
&= (\imod{(4 +3(n - 1))}{n}, \imod{(2* (n - 1) + 2)}{n})\\
& = (1, 0) 
\end{align*}
\begin{theo}
  \label{theo/sn}
For $n > 1$, we have $w^{2^{m-1}} \oplus\, v^{2^{m- 1}} = (\imod{S_m}{n}, 0)$, for $m>1$.
\end{theo}
We prove this by induction.\\
If $m = 1$, we have $w + v = (2, 1) \oplus (2, n -1) = (\imod{4}{n},\imod{n}{n})= (\imod{4}{n}, 0)$.\\
If we suppose that  $w^{2^{m-1}} \oplus\, v^{2^{m- 1}} = (\imod{S_m}{n}, 0)$,
squaring on both side gives
$$(w^{2^{m-1}} \oplus\, v^{2^{m- 1}}) \,\odot\,(w^{2^{m-1}} \oplus\, v^{2^{m- 1}}) = (\imod{S_m}{n}, 0) \,\odot\,(\imod{S_m}{n}, 0)$$
Using the distributivity, commutativity and associativity gives us
$$(w^{2^{m}} \oplus v^{2^{m}}) \oplus (2, 0) \,\odot\, (w^{2^{m-1}}v^{2^{m- 1}}) = (\imod{S_m^2}{n}, 0)$$
For the left side, using some properties of exponentiation we get:
\begin{align*}
(w^{2^{m}} \oplus v^{2^{m}}) \oplus (2, 0) \,\odot\, (w^{2^{m-1}}v^{2^{m- 1}})
& = (w^{2^{m}} \oplus v^{2^{m}}) \oplus (2, 0) \,\odot\, ((wv)^{2^{m-1}}) \\
& = (w^{2^{m}} \oplus v^{2^{m}}) \oplus (2, 0) \,\odot\, (1,0)^{2^{m-1}} \\
& = (w^{2^{m}} \oplus v^{2^{m}}) \oplus (2, 0)
\end{align*}
For the right side, using Definition~\ref{def/sn} we get:
\begin{align*}
(\imod{S_m^2}{n}, 0) = (\imod{(S_{m + 1} + 2)}{n}, 0) = (\imod{S_{m + 1}}{n}, 0) \oplus (2, 0)
\end{align*}
Simplifying by $(2, 0)$ on both side, we get
\begin{align*}
(w^{2^{m}} \oplus v^{2^{m}}) = (\imod{S_{m + 1}}{n}, 0)
\end{align*}

\begin{theo}
  \label{theo/sno}
For $n > 1$ and $m > 1 $, if we have $w^{2^{m-2}} \oplus\, v^{2^{m- 2}} = (0, 0)$, then
$w^{2^{m-1}} \neq (1, 0)$ and $w^{2^{m}} = (1, 0)$.
\end{theo}
Multiplying the left side by $w^{2^{m-2}}$ we get
\begin{align*}
w^{2^{m-2}} \odot (w^{2^{m-2}} \oplus\, v^{2^{m- 2}})
& = (w^{2^{m-2}} \odot w^{2^{m-2}}) \oplus (w^{2^{m-2}} \odot v^{2^{m-2}}) \\
& = w^{2^{m - 1}} \oplus (wv)^{2^{m-2}} \\
& = w^{2^{m - 1}} \oplus {(1, 0)}^{2^{m-2}} \\
& = w^{2^{m - 1}} \oplus {(1, 0)}
\end{align*}
So we get $ w^{2^{m - 1}} = -(1,0) = (n-1, 0)\neq (1,0)$ since $n > 1$. Squaring $ w^{2^{m - 1}} = -(1,0)$
 we get $ w^{2^{m}} = (1,0)$.
\begin{Def}[Mersenne numbers]
  \label{def/merseene}
$M_p$ is the p{\it th} Mersenne if $M_p = 2^p - 1$.
\end{Def}

\begin{theo}[Lucas-Lehmer Test]
  \label{theo/lucas}
If $p > 2$ and $\div{M_p}{S_{p-1}}$ then $M_p$ is prime.
\end{theo}
The proof is by contradiction. We suppose that $M_p$ is composite, so there exists an $n$ such that $1 < n \le \sqrt{M_p}$ and $\div{n}{M_p}$. 
We consider $K_n^*=I(K_n)$. As $(0, 0) \not\in K_n^*$, we have $\size{K_n^*} \le n^2 - 1 < M_p$. 
By Theorem~\ref{theo/wv}, we have $w \odot v = (1, 0)$, so $w\in K_n^*$. We have $\div{n}{M_p}$ and $\div{M_p}{S_{p-1}}$,
so $\imod{S_{p-1}}{n} = 0$.
By Theorem~\ref{theo/sn}, we get $w^{2^{m-2}} \oplus v^{2^{m-2}} = (0, 0)$. By Theorem~\ref{theo/sno}, we
deduce that $w^{2^{m-1}} \neq (1, 0)$ and $w^{2^{m}} = (1, 0)$. 
By Theorem~\ref{theo/powdiv}, we have $\div{o(w)}{2^m}$. We deduce that $o(w)= 2 ^p$ for some
$p \le m$. If $p < m$, we would have $w^{2^{m-1}} = w^{2 ^{p + (m-1 - p)}} = (w^{2^p})^{2^{m-1 -p}} = (1,0)$, so
$o(w) = 2^p$. But by Theorem~\ref{theo/order}, we have $\div{o(a)}{\size{K_n^*}}$, so in particular we have
$2^p \le \size{K_n^*}$. Putting everything together, we get a contradiction $2^p \le \size{K_n^*} \le n^2 - 1 < M_p = 2^p -1$.

\section{Pocklington certificate \label{pock}}

Pocklington certificate let us assess the primality of a number $n$ by collecting enough factors of $n-1$ and showing 
that these factors verify a given relation. 
\begin{theo}[Pocklington]
  \label{theo/pockl}
If $F_1 > 1$, $R_1 > 0$ and $N - 1 = F_1  R_1$, if we have that for each prime number $p$ such that $\div{p}{F_1}$
there exists an $a$ such that $\mod{a^{N-1}}{1}{N}$ and $\gcd{(a ^{(N - 1)/ p} -1)}{N}=1$, then 
for each prime $n$ such that $\div{n}{N}$ we have $\mod{n}{1}{F_1}$.
\end{theo}
To prove $\mod{n}{1}{F_1}$, we show that $\div{F_1}{n-1}$. It is enough to prove that $\div{p ^\alpha}{n-1}$ 
for each prime number $p$ such that $\div{p^\alpha}{F_1}$. If $\alpha \ge 1$,
we have $\div{p}{F_1}$. So there exists $a$ such that $\mod{a^{N-1}}{1}{N}$ and $\gcd{(a ^{(N - 1)/ p} -1)}{N}=1$.
We have $\imod{a}{n} \in \zn{n}$, so $o(\imod{a}{n})$ makes sense. We are going to prove that $\div{p^\alpha}{o(\imod{a}{n})}$
which is enough since $\div{o(\imod{a}{n})}{n-1}$ by Theorem~\ref{theo/orderzp}.
We have $\div{n}{N}$, so $\mod{a^{N-1}}{1}{N}$ implies $\mod{(\imod{a}{n})^{N-1}}{1}{n}$.
By Theorem~\ref{theo/powdiv} we get $\div{o(\imod{a}{n})}{N-1}$. We have also $\gcd{(a ^{(N - 1)/ p} -1)}{N}=1$, so
in particular as $\div{n}{N}$ we have $\gcd{(a ^{(N - 1)/ p} -1)}{n}=1$. If we had $\div{o(\imod{a}{n})}{(N-1)/p}$,
by Theorem~\ref{theo/powdiv} we would get that $\mod{a ^{(N - 1)/ p}}{1}{n}$, so we would have $n = \gcd{(a ^{(N - 1)/ p} -1)}{n}$.
To sum up, we have that $\div{o(\imod{a}{n})}{N-1}$ but also that ${o(\imod{a}{n})}\not\!|\, {(N-1)/p}$. This means that ${o(\imod{a}{n})}$ contains
all the power of $p$, i.e. for all $\beta$ such that $\div{p^\beta}{N-1}$ we have $\div{p^\beta}{o(\imod{a}{n})}$. 
So we get that $\div{p^\alpha}{o(\imod{a}{n})}$.

\begin{theo}
  \label{theo/pocklcorr}
If $F_1 > 1$, $\div{F_1}{N-1}$ and $F_1 > N$, if we have that for each prime number $p$ such that $\div{p}{F_1}$
there exists an $a$ such that $\mod{a^{N-1}}{1}{N}$ and $\gcd{(a ^{(N - 1)/ p} -1)}{N}=1$, then 
$N$ is prime.
\end{theo}
This is a direct corollary of Theorem~\ref{theo/pockl}. If $N$ was composite, there would be an $n$ such that $1 < n \le \sqrt{N}$
and $\div{n}{N}$. We have $n \le \sqrt{N} < F_1$ and by Theorem~\ref{theo/pockl} we also have $\mod{n}{1}{F_1}$. So we deduce
that $n = 1$ which contradicts $1 < n$.

Now we can derive two usual tests from this last corollary.
\begin{Def}[Fermat numbers]
  \label{def/fermatn}
$F_p$ is the p{\it th} Fermat number if $F_p = 2^{2^p} - 1$.
\end{Def}

\begin{theo}[Pepin Test]
  \label{theo/pepin}
If $p > 1$ and $\mod{3 ^{(F_p - 1)/ 2}}{-1}{F_p}$ then $F_p$ is prime.
\end{theo}
This is a direct application of Theorem~\ref{theo/pocklcorr} with $a=3$ and $F_1 = 2 ^{2 ^ p}$.
\begin{theo}[Proth Test]
  \label{theo/proth}
If $p = h 2 ^k + 1$ with $2^k > h$ and there exists an $a$ such that $\mod{a^{(n - 1)/ 2}}{-1}{p}$ then $p$ is prime.
\end{theo}
This is a direct application of Theorem~\ref{theo/pocklcorr} with $F_1 = {2 ^ k}$.

Theorem~\ref{theo/pocklcorr} requires to  be able to factorize $N-1$ till $\sqrt{N}$. We can do considerably better
($\sqrt[3]{N/2}$ instead of $\sqrt{N}$) with the following theorem.

\begin{theo}
  \label{theo/pocklextra}
Let $F_1 > 1$, $R_1 > 0$ and $N-1 = F_1 R_1$, such that $F_1$ is even and $R_1$ is odd,
let $m \ge 1$ , $s := R_1 / (2 F_1)$, $r = \imod{R_1}{(2 F_1)}$  such that $N < (m F_1 + 1) * (2 F_1^2  + (r - m) F_1 + 1)$
and for all $\lambda$ such that $1 \le \lambda < m$, we have ${(\lambda F_1 + 1)}{\not\!|}\, N$,
if for each prime number $p$ such that $\div{p}{F_1}$ there exists an $a$ such 
that $\mod{a^{N-1}}{1}{N}$ and $\gcd{(a ^{(N - 1)/ p} -1)}{N}=1$, 
 then if $s = 0$ or $r^2 - 8s$ is not a square then $N$ is prime.
\end{theo}
We proceed by contradiction. Suppose that $N$ is composite, we are going to prove that $s \neq 0$ and $r² - 8s$ is a square.
$N$ is composite, so there exist $K_1$, $K_2$ such that $N=K_1K_2$. By Theorem~\ref{theo/pockl} we know that
$\mod{K_i}{1}{F_1}$. So there exist $c$, $d$ such that $N = (c F_1 + 1) (d F_1 +1)$. Furthermore the fact that 
for all $\lambda$ such that $1 \le \lambda < m$, we have ${(\lambda F_1 + 1)}{\not\!|}\, N$ gives us 
that $c \ge m$ and $d \ge m$. 
We have $N-1 = F_1 R_1$ and $N = cd F_1^2 + (c + d)F_1 + 1$ so $R_1 = cd F_1 + (c + d)$.
We have also that $R_1$ is odd and $F_1$ is even, so $cd F_1$ is even which implies that $c + d$ is
odd. So $cd$ must be even. We have $R_1 = (cd/2)2F_1 + (c + d)$ and by definition of $s$ and $r$
we have also $R_1 = s2F_1 + r$. If we manage to prove that $c + d = r$ we are done since $s = (cd/2) \neq 0$
and $r^2 -8 s = {(c + d)}^2 - 4cd = {(c - d)}^2$.

To prove $c + d = r$, as we have $(cd/2)2F_1 + (c + d) = s2F_1 + r$ and $r = \imod{R_1}{(2 F_1)}$ 
we  know that $r = \imod{c + d}{(2 F_1)}$ , we then just need to prove that $(c + d) - r < 2F_1$ to conclude.
We have
\begin{align*}
(m F_1 + 1) * (2 F_1^2  + (r - m) F_1 + 1) > N = cd F_1^2 + (c + d)F_1 + 1
\end{align*}
We have $(c -m)(d - m) \ge 0$, so $cd \ge m(c + d) - m^2$. Using this inequality to minor the right
side of the previous equation we get:
\begin{align*}
(m F_1 + 1) * (2 F_1^2  + (r - m) F_1 + 1)  & >(m(c+ d) - m^2) F_1^2 + (c + d)F_1 + 1\\
& = (mF_1 + 1)(((c + d) - m) F_1 + 1)
\end{align*}
Simplifying we get $2 F_1^2  + (r - m) F_1 + 1  > ((c + d) - m) F_1 + 1$ so $(c + d) - r < 2F_1$.

\section{Acknowledgments}

This formalisation has been mainly motivated by Benjamin Gregoire's quest
for large prime numbers verified by {\sc Coq}.

Proofs of Sections~\ref{lucas} and ~\ref{pock} are transcriptions
from the beautiful site \url{http://primes.utm.edu/prove}. 
Proofs of  Theorems~\ref{theo/pocklextra}
and~\ref{theo/pockl} are transcriptions of the seminal paper~\cite{lehmer}.

\bibliographystyle{plain} 
\bibliography{Note}

\end{document}